\documentclass[conference]{IEEEtran}
\ifCLASSINFOpdf
  \usepackage[pdftex]{graphicx}
\else
\fi
\usepackage{amsmath}
\usepackage{amsfonts}
\usepackage{multirow}
\usepackage{algorithm}
\usepackage[noend]{algpseudocode}

\IEEEoverridecommandlockouts

\begin{document}
%
\title{Graph based Platform for Electricity Market Study, Education and Training}
%
%
%
\author{\IEEEauthorblockN{Tao Chen, Chen Yuan, Guangyi Liu, Renchang Dai} 
\IEEEauthorblockA{GEIRI North America, San Jose, CA, USA \\
Email: guangyi.liu@geirina.net}
\thanks{This work was supported by State Grid Corporation of China technology project SGRIJSKJ (2016)800.}
}

\maketitle

\begin{abstract}
With the further development of deregulated electricity market in many other countries around the world, a lot of challenges have been identified for market data management, network topology processing and fast market-clearance mechanism design. In this paper, a graph computing framework based on TigerGraph database is proposed to solve a security constrained unit commitment (SCUC) and security constrained economic dispatch (SCED) problem, with parallelized graph power flow (PGPF) and innovative LU decomposition techniques, for electricity market-clearance. It also provides a comprehensive visualization platform to demonstrate the market clearing results vividly, such as locational marginal price (LMP), and is able to be utilized for electricity market operators' education and training purpose.
\end{abstract}


%
\IEEEpeerreviewmaketitle

\section{Introduction}
%
%
%
%
\IEEEPARstart{W}{ith} the increasing complexity in large-scale power systems and further development of deregulated electricity market in many countries, a lot of challenges have been identified for electricity market data management, power system operation, network topology processing and fast market-clearance mechanism design. Especially under the competitive environment of power industry, bidding and clearing process based on nodal price theory has been thoroughly studied by academia and a number of regional transmission organization (RTO) or independent system operators (ISOs), such as PJM, New York ISO and ISO New England. The two-settlement system involving day-ahead market (DAM) and real-time markets (RTM) is most widely adopted for considering economic efficiency and effectiveness with various type market participants. Theoretically, the market operator for the two-settlement system will solve a bid/offer-based security-constraint unit commitment (SCUC) and security constraint economic dispatch (SCED) problem to determine the pricing solution and dispatch results. The calculation process involves many market information submitted by participants, such as the characteristics of generating units, transmission capacity, generation offers and demand bids, scheduled transactions, curtailment contracts, and so on \cite{scucAC}.
\\ \indent Many simulation platforms are developed for modeling this market-clearance process through time over an AC transmission grid with congestion managed by locational marginal pricing (LMP). For instance, in \cite{AMES}, an AMES Wholesale Power Market Test Bed with 8-Zone ISO-NE Test System permits users to vary the generation mix in wholesale markets through a stochastic SCUC formulation. Some researchers are in collaboration with industry partners to make use of ISO/RTO-scale systems for more realistic simulation \cite{RTO}. Many other commercial electricity market simulation platforms, such as PLEXOS \cite{plexos} and EMCAS \cite{emcas}, are also widely adopted and studied for supporting decision making and market participants' training. In the development of these simulation platforms, SCUC and SCED models are believed to be the fundamental for clearing electricity market either in day-ahead or real-time scenario \cite{exPost}. That's why there are so many discussion about unit commitment (UC) models and its coupling problem with power flow constraints \cite{scucAC}\cite{scucAnci}\cite{wulei}. Many UC or SCUC models embedded in electricity market simulation framework prefer to use DC power flow model because of its computational efficiency, stability and capability to avoid time-consuming iteration process caused by nonlinear AC power flow constraints \cite{zechun}. Indeed, the results of the DC approximation are close to the full AC solutions, especially with some modification, like fictitious nodal demand (FND) model \cite{DCOPF}, to offset the effect of active power losses. However, there is still a great desire to combine the AC power flow model with SCUC problem in a computationally efficient way, which benefits economic operation of power system and electricity pricing in market.
\\ \indent In this paper, a graph computing framework based on TigerGraph \cite{TigerGraph} is proposed to facilitate the electricity market clearing process by solving SCUC and SCED problems in a visualized way, with help of parallelized graph power flow (PGPF) and innovative LU decomposition techniques. The contribution is claimed mainly by providing many flexible interfaces for market or network data storage/operation and a comprehensive visualization platform to demonstrate the electricity market clearing results with the utilization of graph computing for considering power network constraints. This TigerGraph based simulation platform for electricity market can also contributes to helping electricity market operators' education and training, thus promoting market deregulation.




\section{Graph computing framework}

\subsection{Graph computing and power flow with TigerGraph}

\begin{figure}[!htb]
\centering
\includegraphics[width=0.40\textwidth]{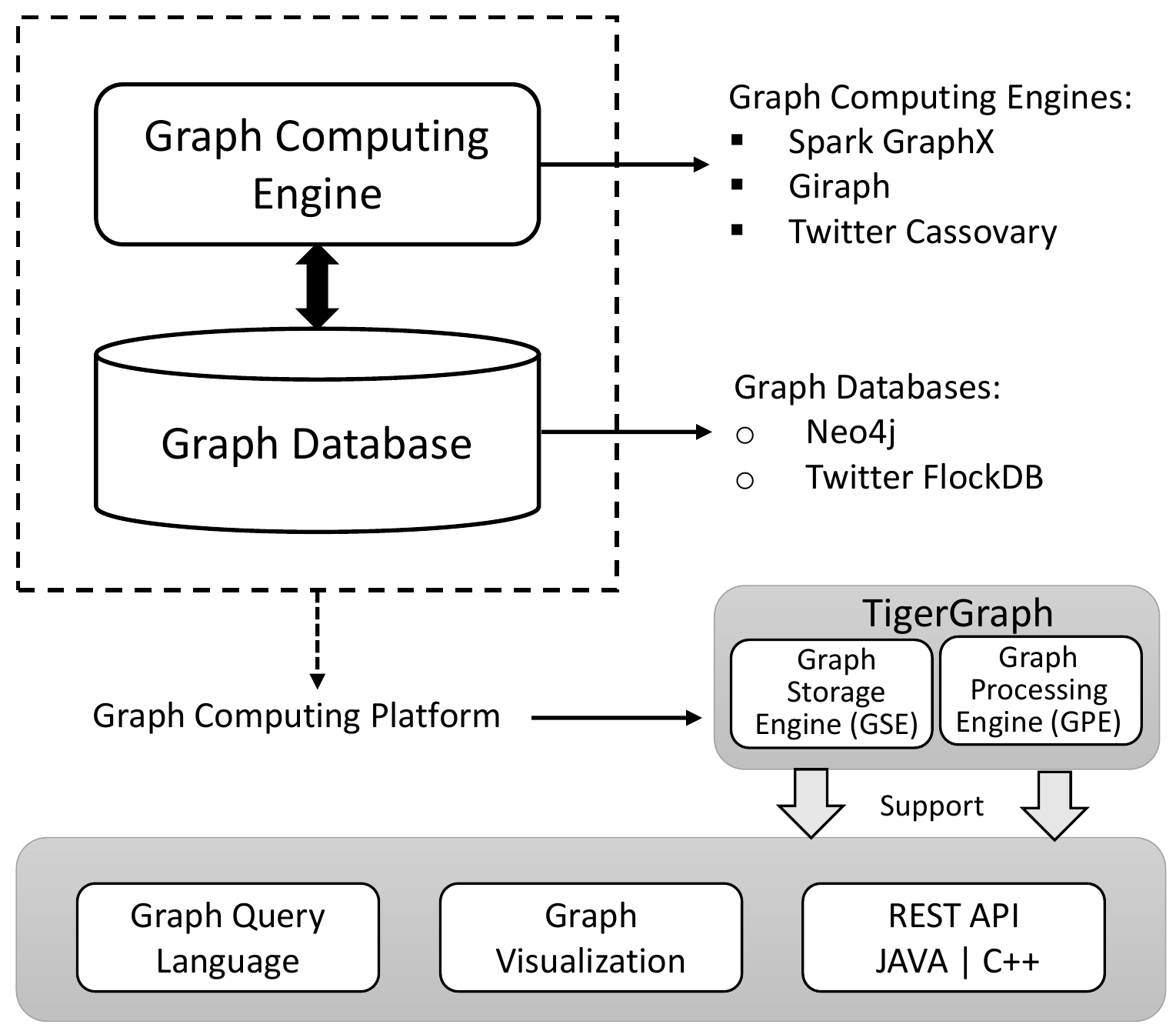}
\caption{TigerGraph with its main blocks}
\label{gsql}
\end{figure}

Graph computing is widely used in social networks, Internet searches, biological gene maps, intelligent transportation and so on, which usually involves hundred millions scale data analysis or computing. Graph computing, consisting of the structure (graph) and the process (traversal), is becoming a promising supplement of the traditional computing in solving the large scale data management and complicated computation technical challenges. The graph structure is the data model defined using a vertex/edge/property topology. Graph computing can be easily implemented using the graph data structure, in which each vertex or edge can have its own computation function or logic. Therefore, vertices and edge are not only data storage units but also dynamic computation units. Communication between vertices in a graph structure is realized by exchanging information through edges. "Local computation" and "coordinated collaboration" are used in a graph computing so that large-scale parallel computing can be easily achieved. In a word, graph computing are especially useful for complex system, such as power system, which usually require multiple iterations for calculation \cite{liu}.
\\ \indent TigerGraph doesn't work only as database for data storage but also work as computing engine that enables parallel computing mechanism. It combines different features of other graph databases and graph computing engines, with supporting external applications (Figure \ref{gsql}). TigerGraph unifies the map-reduce and parallel graph processing based on bulk synchronous parallel (BSP) model \cite{bsp}.

\begin{figure*}[!htb]
\centering
\includegraphics[width=0.75\textwidth]{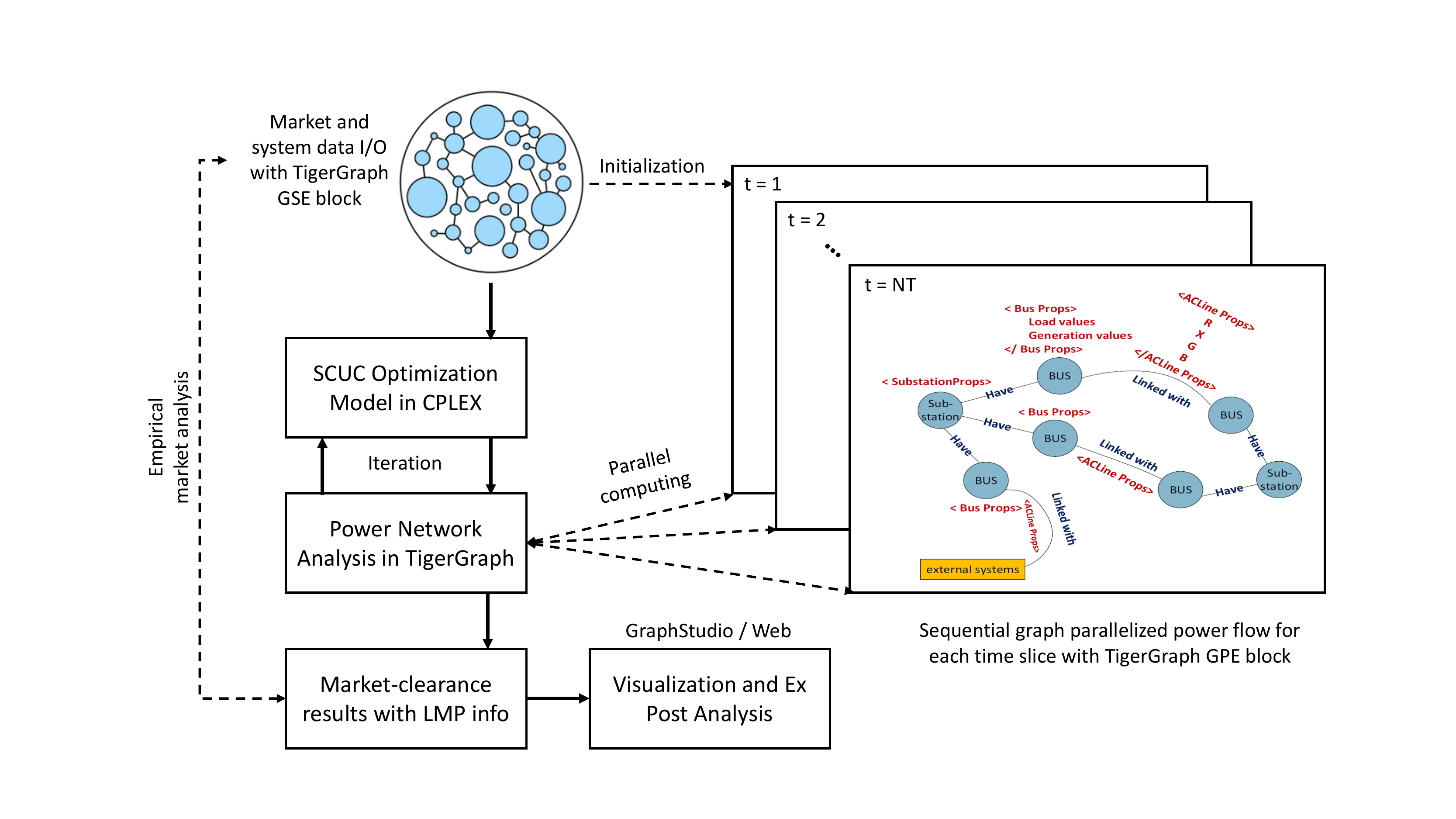}
\caption{Graph computing framework for electricity market based on TigerGraph}
\label{frame}
\end{figure*}

The conventional topology processing and power flow analysis strongly relies on indexing, needing two separate data sheets for branch (i.e. transmission line) and node (i.e. bus) information, which adversely impact the performance of data storage and computational efficiency. However, the graph-based framework and particular algorithm allow the POST-ACCUM statement to use BSP graph parallel calculation \cite{bsp} for fast solving unit commitment and power flow problems, as shown in Figure \ref{frame} and Algorithm \ref{al}.

\makeatletter
\def\BState{\State\hskip-\ALG@thistlm}
\makeatother

\begin{algorithm}
\caption{Graph Parallelized Power Flow}\label{euclid}
\begin{algorithmic}[1]
\Procedure{Fast Decoupled Power Flow}{}
\State $\mathbf{V_s}, \mathbf{V_r}, \mathbf{V_m} \gets \text{Info from }\textit{VERTEX}$
\State $\mathbf{B^{\prime}}, \mathbf{B^{\prime \prime}}, \textit{Insert\_LU} \gets \text{Info from }\textit{EDGE}$
\BState \emph{top}:
\If {$\textit{Insert\_LU} = \text{NULL}$} layer segmentation
\State $\text{LU factorization with CSR}$
\EndIf
\State $\mathbf{\Delta P}, \mathbf{\Delta Q} \gets \textit{Initial value}$
\BState \emph{loop}:
\If {$\textit{max} \{ | \mathbf{\Delta P} + j \mathbf{\Delta Q} | \} \leq \epsilon $}
\State $\text{traversal process for } \textit{EDGE}$.
\State $\text{traversal process for } \textit{VERTEX}$.
\State $k \gets k+1$.
\State \textbf{goto} \emph{loop}.
\State \textbf{close};
\EndIf
\State \Return $\textit{AccumList}\{ \mathbf{P}_l, \mathbf{P}_{loss} \}$
\EndProcedure
\end{algorithmic}
\label{al}
\end{algorithm}


\subsection{SCUC and SCED}

Some reserve requirements are also considered in this paper, which include regulation, spinning reserve, 10-minute non-spinning reserve and 30-minute operating reserve. An important aspect of reserve constraints is their hierarchical nature that allows the substitution of a higher quality service with a lower quality one. Faster response reserves are graded as higher quality \cite{reserve}. Considering both the system reserve and energy-only requirement, the objective function of a SCUC model can be formulated as (\ref{obj}), with $F_{c,i}(P_{i,t,j})$ indicating stepwise bidding cost function; $SU_{i,t}$ ($SD_{i,t}$) for start-up (shut-down) cost of generator $i$ at time $t$ with binary indicator $W_{i,t}$ ($Y_{i,t}$); and $F_{\cdot,i}(R_{i,t}^\cdot)$ for various types of reserve of generator $i$ at time $t$.
\begin{equation}
\begin{split}
\label{obj}
min \quad \sum_{i \in \mathcal{G}} \sum_{t=1}^{NT} \bigg [ F_{c,i}(P_{i,t,j}) + SU_{i,t}  W_{i,t}+ SD_{i,t}  Y_{i,t}  \\
+ F_{r,i}(R_{i,t}^r) + F_{sp,i}(R_{i,t}^{sp}) + F_{n1,i}(R_{i,t}^{n1}) + F_{n3,i}(R_{i,t}^{n3}) \bigg] 
\end{split}
\end{equation}
s.t.
\begin{equation}
\label{ba}
\sum_{i \in \mathcal{G}} \sum_{j=1}^{NL} P_{i,t,j} = D_t,    \quad  t = 1,2, ..., NT
\end{equation}
\begin{equation}
\label{con1}
I_{i,t} - I_{i,t-1} - W_{i,t} + Y_{i,t} =0,  \quad  t=2, ..., NT,  \quad  \forall i \in \mathcal{G}
\end{equation}
\begin{equation}
\label{con2}
W_{i,t} + Y_{i,t} \leq 1,  \quad  t=1, 2, ..., NT,  \quad  \forall i \in \mathcal{G}
\end{equation}
\begin{equation}
\begin{split}
\label{con22}
\Bigg [ \text{min} \Big ( NT-t, T_i^{up}-1 \Big ) \Bigg ] W_{i,t} \leq \sum_t^{t + \text{min}(NT-t, T_i^{up}-1)} I_{i,t} \\
 \quad t = 1,2, ..., NT-1, \quad  \forall i \in \mathcal{G}
\end{split}
\end{equation}
\begin{equation}
\begin{split}
\label{con222}
\Bigg [ \text{min} \Big ( NT-t, T_i^{dn}-1 \Big ) \Bigg ] Y_{i,t} + \sum_t^{t + \text{min}(NT-t, T_i^{dn}-1)} I_{i,t} \\ 
\leq  \text{min} \Big ( NT-t, T_i^{dn}-1 \Big ), \quad t = 1,2, ..., NT-1, \quad  \forall i \in \mathcal{G}
\end{split}
\end{equation}
\begin{equation}
\label{con2222}
\sum_{t=1}^{NT} W_{i,t} \leq NS_i, \quad  \forall i \in \mathcal{G}
\end{equation}
\begin{equation}
\label{ini}
I_{0,i} = I_{i,Initial},    \quad  \forall i \in \mathcal{G}
\end{equation}
\begin{equation}
\begin{split}
\label{con3}
P_{i,t,j} - P_{i,j}^{UB} \times I_{i,t} \leq 0, \quad  t = 1,2, ..., NT, \quad  \forall i \in \mathcal{G} \\
\quad j = 1,2, ..., NL
\end{split}
\end{equation}
\begin{equation}
\begin{split}
\label{con4}
P_{i,t,j} \geq 0, \quad  t = 1,2, ..., NT, \quad  \forall i \in \mathcal{G} \\
 \quad j = 1,2, ..., NL
\end{split}
\end{equation}
\begin{equation}
\begin{split}
\label{trans}
\sum_{i \in \mathcal{G}} \bigg [ GSF_{i,l} \times \Big ( \sum_{j=1}^{NL} P_{i,t,j}^{k} - \sum_{j=1}^{NL} P_{i,t,j}^{k-1} \Big ) \bigg ]  \leq P_{l}^{UB} - P_{l}^{k-1}, \\
\quad  t = 1,2, ..., NT, \quad  \forall l \in \mathcal{L}
\end{split}
\end{equation}
\begin{equation}
\begin{split}
\label{res}
0 \leq R_{i,t}^{r} \leq R_{max,i}^{r} I_{i,t}, \quad 0 \leq R_{i,t}^{sp} \leq R_{max,i}^{sp} I_{i,t},  \\
0 \leq R_{i,t}^{n1} \leq R_{max,i}^{n1}, \quad 0 \leq R_{i,t}^{n3} \leq R_{max,i}^{n3}, \\
t = 1,2, ..., NT, \quad  \forall i \in \mathcal{G}
\end{split}
\end{equation}
\begin{equation}
\label{r}
\sum_{i \in \mathcal{G}} R_{i,t}^r \geq R_{t}^r,    \quad  t = 1,2, ..., NT
\end{equation}
\begin{equation}
\label{sp}
\sum_{i \in \mathcal{G}} R_{i,t}^{sp} \geq R_{t}^{sp},    \quad  t = 1,2, ..., NT
\end{equation}
\begin{equation}
\label{n1}
\sum_{i \in \mathcal{G}} \Big (R_{i,t}^{sp} + R_{i,t}^{n1} \Big ) \geq R_{t}^{sp} + R_{t}^{n1},    \quad  t = 1,2, ..., NT
\end{equation}
\begin{equation}
\begin{split}
\label{n3}
\sum_{i \in \mathcal{G}} \Big (R_{i,t}^{sp} + R_{i,t}^{n1} + R_{i,t}^{n3} \Big ) \geq R_{t}^{sp} + R_{t}^{n1} + R_{t}^{n3}, \\
t = 1,2, ..., NT
\end{split}
\end{equation}
\begin{equation}
\begin{split}
\label{resl}
R_{i,t}^{r} + R_{i,t}^{sp} + R_{i,t}^{n1} + R_{i,t}^{n3} + \sum_{j=1}^{NL} P_{i,t,j} \leq P_{i}^{MAX}, \\
t = 1,2, ..., NT, \quad  \forall i \in \mathcal{G}
\end{split}
\end{equation}
\begin{equation}
\begin{split}
\label{con5}
I_{i,t}, W_{i,t}, Y_{i,t} \in \{ 0,1 \}, \quad  t = 1,2, ..., NT, \quad  \forall i \in \mathcal{G}
\end{split}
\end{equation}

In constraints, eq. (\ref{ba}) ensures system balance between supply and demand, and eq. (\ref{con1})-(\ref{ini}) guarantee that start-up and shunt-down characteristics, including minimum up (\ref{con22}) and down (\ref{con222}) time limits, are satisfied with generator turn-on indicator, $I_{i,t}$. The consideration for ramp rate limit and output lower bound are given in eq. (\ref{con3}) and  (\ref{con4}). Eq. (\ref{trans}) provides transmission line capacity limit constraint with generation shift factor, $GSF_{i,l}$, of generator $i$ for transmission line $l$. It's worthy mentioning that the generator output interval $P_{i,t,j}$ modeled and calculated in CPLEX will go though some iteration process with line power flow $P_{l}$ information collected from AC power flow calculation within TigerGraph. The subscript $k-1$ and $k$ indicates the $(k-1)$th and $k$th iteration respectively. The hierarchical constraints of various types of reserves from high quality to low quality are given in (\ref{res})-(\ref{n3}). Eq. (\ref{resl}) guarantees the maximum limit of generator output is not violated, and (\ref{con5}) explicitly indicates all the binary variable constraints.
\\ \indent Finally, the spot electricity prices are calculated in the real-time market in the form of either ex-ante for example NY ISO, or ex post, for instance ISO New England, PJM, and Midwest ISO \cite{exPost}. Thus, a standard SCED problem is proposed and solved via ex-ante formulation (\ref{objED})-(\ref{con2ED}), in which, the generation dispatches and LMPs are obtained from the same optimization model.  
\begin{equation}
\label{objED}
min \quad \sum_{i \in \mathcal{G}} C_i \Big (   P_{i}  \Big ) 
\end{equation}
s.t.
\begin{equation}
\label{baED}
\sum_{i \in \mathcal{G}} P_{i} \times DF_i - \sum_{h \in \mathcal{D}} D_h \times DF_h - P_{loss} (F_l) = 0
\end{equation}
\begin{equation}
\label{con1ED}
P_{i}^{min} \leq P_{i} \leq P_{i}^{max}, \quad \forall i \in \mathcal{G}
\end{equation}
\begin{equation}
\label{con2ED}
-F_{l}^{max} \leq F_{l} \leq F_{l}^{max}, \quad \forall l \in \mathcal{L}
\end{equation}

\subsection{LMP with visualization}
The electricity pricing information at different nodes (i.e. LMP) can be obtained by solving SCED problem, and consists of different price components. The LMP at bus $i$, $LMP_i$, is not dependent on the slack bus (reference), however the split into LMP components, $LMP_i^{cong}$ and $LMP_i^{loss}$, is dependent on the selection of the slack bus, due to the calculation of generation shift factor, $GSF_{i,l}$. The $DF_i$ in (\ref{lmpl}) is a delivery factor, which shows how much power is going to reach the reference bus if additional $1$ MW is injected at bus $i$.
\begin{equation}
\label{lmp}
LMP_i = LMP^{fuel} + LMP_i^{cong} + LMP_i^{loss}
\end{equation}
\begin{equation}
\label{lmpc}
LMP_i^{cong} = \sum_{l \in \mathcal{L}} GSF_{i,l} \times \mu_l
\end{equation}
\begin{equation}
\label{lmpl}
LMP_i^{loss} = c_i^{fuel} \times \Big (DF_i - 1 \Big )
\end{equation}

Data visualization has become a rapidly evolving blend of science and art that promotes reviving the traditional industry \cite{vi}. Visualization of power network and electricity market operation can mimic the decision-making process for determining nodal price with various system effects, such as transmission congestion, topology change and marginal generation limit. It helps understanding market power in electric power systems \cite{vipower} and provides some tools for training or educating potential practitioners in electricity market.

\section{Simulation and test case}
The significant advantages of the graph based framework are the flexibility for data management and the scalability for large-scale system. For instance, we can store various kinds of parameters, especially temporal information (e.g. bidding interval, LMP results, load profiles, etc.), in the data structure of time-tree edges for request of intuitive observation. The network with hundreds and even thousands nodes can also be efficiently analyzed by calling graph parallel implementation. These features can be shown in following test cases. And all the test cases are developed and implemented in C/C++ with CPLEX and TigerGraph on a Linux server, which has 96 2.10GHz Intel Xeon E7 processors and 1TB of global shared memory.

\subsection{IEEE-14 bus system}

In this test case, an additional generator is added at node No. 3 (Figure \ref{14bus}) with its bidding curve parameters, along with all the other generators, given in Table \ref{bid}. The transmission line between node No. 2 and node No. 3 is assumed as a congested line with limited capacity 0.40 MWA. The results are given in Table \ref{result} and Figure \ref{14bus} with running time 2804.396 $ms$ and 8 threads.

\begin{figure}[!htb]
\centering
\includegraphics[width=0.4\textwidth]{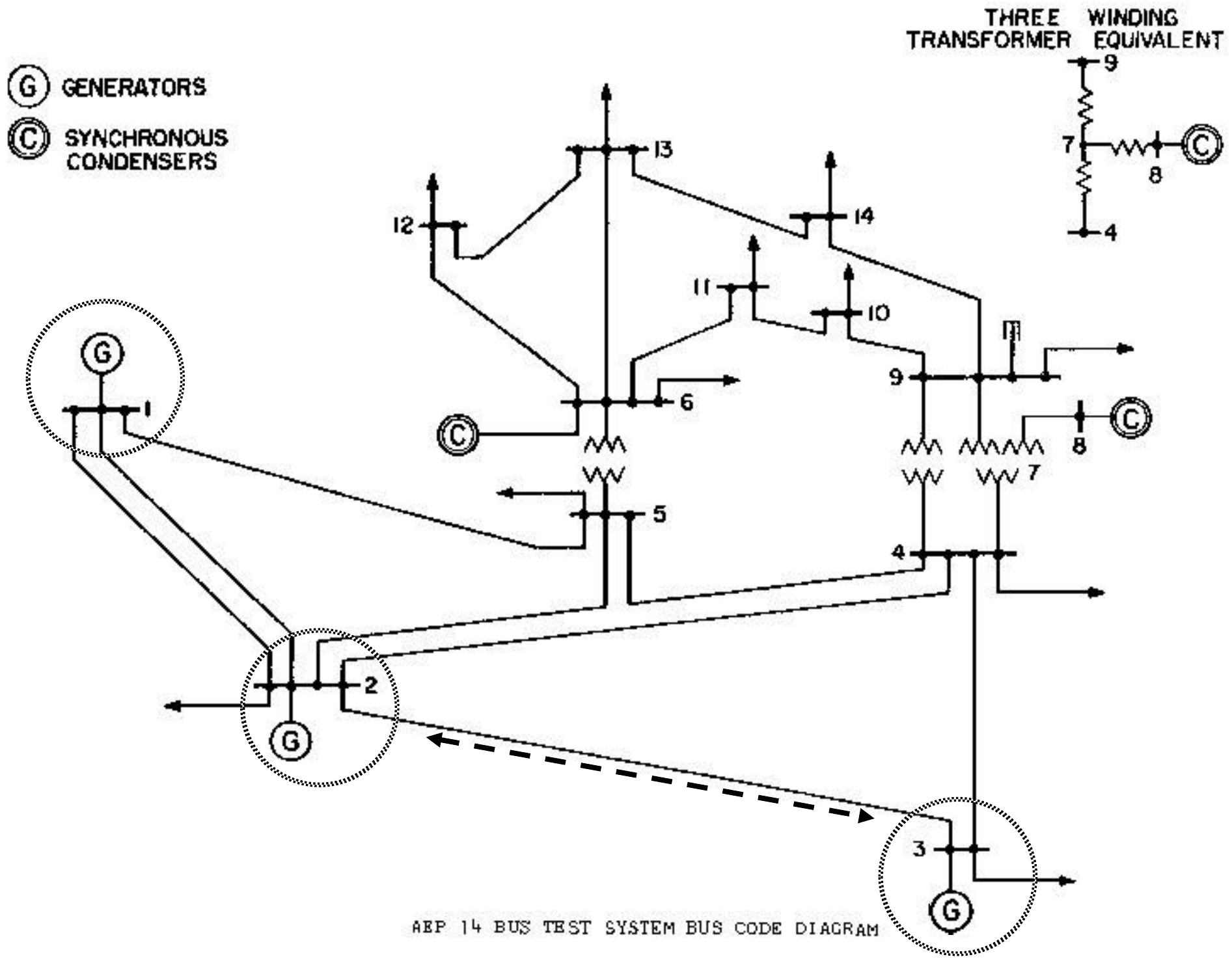}
\caption{Modified IEEE-14 bus system}
\label{14bus}
\end{figure}

\begin{table}[!h]
\renewcommand{\arraystretch}{1.4}
\caption{Bidding price $B$ (\$/MW)  and amount interval $M$ (MW)}
\label{bid}
\centering
\begin{tabular}{c|cccccc}
\hline
 Generator &  $B_1$          &  $M_1$  &  $B_2$           &  $M_2$  &  $B_3$           &  $M_3$ \\
\hline
\hline
No. 1      & 24.1  & 60.0 & 26.9 & 50.0  & 29.7 & 40.0  \\
\hline
No. 2     & 26.7  & 70.0 & 29.6 & 60.0  & 32.7 & 50.0  \\
\hline
No. 3     & 35.7  & 65.0 & 42.6 & 55.0  & 51.1 & 35.0  \\
\hline
\end{tabular}
\end{table}

\begin{table}[!h]
\renewcommand{\arraystretch}{1.1}
\caption{Result from TigerGraph simulation (\$/MW)}
\label{result}
\centering
\begin{tabular}{c|ccccccc}
\hline
 Hour &  $I_{g1}$  &  $I_{g2}$  &  $I_{g3}$ &  $P_{g1}$  &  $P_{g2}$  &  $P_{g3}$  & $LMP_{ave}$ \\
\hline
1       &1 &1 & 0 & 1.10 & 0.35 & 0 & 25.81 \\
2      &1  &1 & 0 & 1.05  & 0.35 & 0 & 26.29 \\
3      &1  &1 & 0 & 0.95 & 0.35 & 0 & 25.26 \\
4      &1  &1 & 0 & 1.05  &0.35 & 0 & 26.02 \\
5      &1  &1 & 0 & 0.95  & 0.35 & 0 & 26.28 \\
6       &1  &1 & 0 & 1.05  &0.35 & 0 & 26.23 \\
7       &1  &1 & 0 & 1.15  & 0.35 & 0 & 27.32 \\
8      &1  &1 & 0 & 1.05  & 0.35 & 0 & 26.20 \\
9      &1  &1 & 0 & 0.95  & 0.35 & 0 & 25.92 \\
10       &1  &1 & 0 & 1.05  & 0.35 & 0 & 26.70 \\
11       &1  &1 & 1 & 1.15  & 0.35 & 0.05 & 28.32 \\
12      &1  &1 & 0 & 1.05  & 0.35 & 0 & 26.12 \\
13      &1  &1 & 0 & 0.95  & 0.35 & 0 & 25.78 \\
14       &1  &1 & 0 & 1.05  & 0.35 & 0 & 26.19 \\
15      &1  &1 & 0 & 1.15  & 0.35 & 0 & 26.10 \\
16       &1  &1 & 0 & 1.05  & 0.35 & 0 & 26.78 \\
17      &1  &1 & 0 & 0.95  & 0.35 & 0 & 25.30 \\
18      &1  &1 &0 & 1.05  & 0.35 & 0 & 26.10 \\
19      &1  &1 &0 & 1.15  & 0.35 & 0 & 26.98 \\
20       &1  &1 & 0 & 0.95  & 0.35 & 0 & 25.82 \\
21       &1  &1 & 0 & 1.05  & 0.35 & 0 & 26.60 \\
22      &1  & 1 & 0 & 1.15  & 0.35 & 0 & 26.82 \\
23      &1  &1 & 0 & 0.95  & 0.35 & 0 & 25.91 \\
24      &1  & 1 & 0 & 1.05  & 0.35 & 0 & 26.75 \\
\hline
\end{tabular}
\end{table}

\begin{figure}[!htb]
\centering
\includegraphics[width=0.48\textwidth]{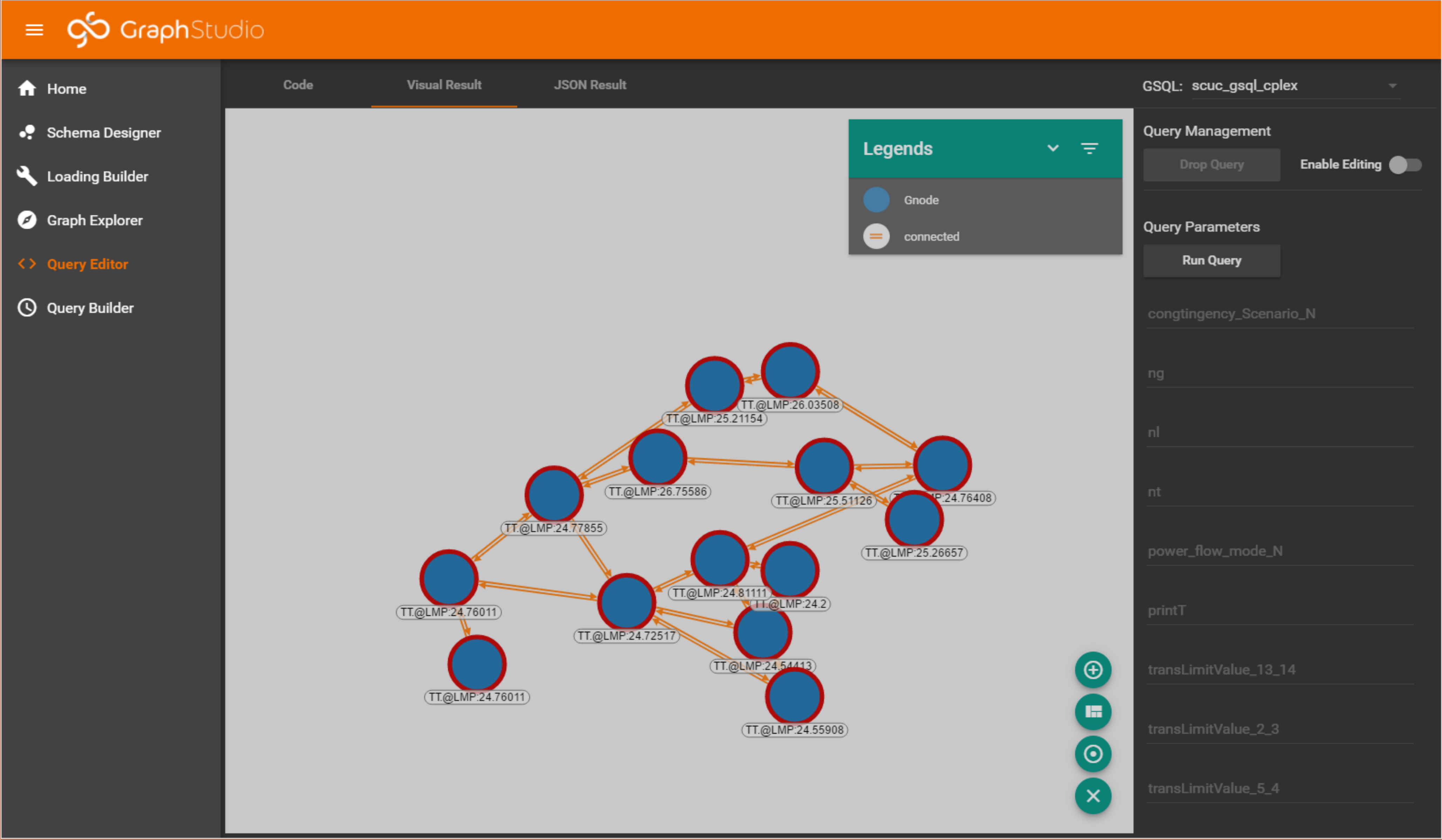}
\caption{Illustrative 14 bus example with visualization on GraphStudio}
\label{14bus}
\end{figure}

\begin{figure}[!htb]
\centering
\includegraphics[width=0.43\textwidth]{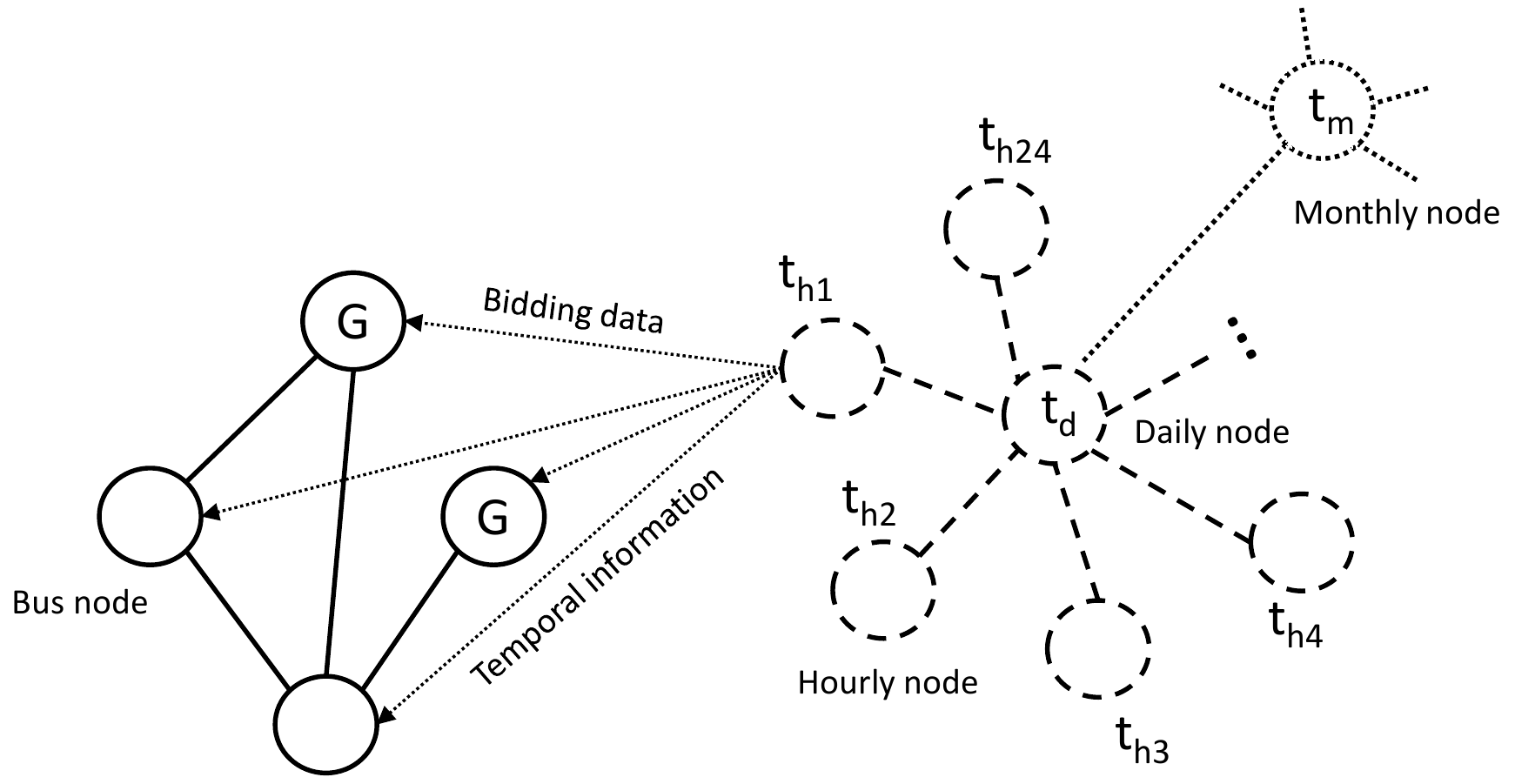}
\caption{Time tree structure in graph database}
\label{timeTree}
\end{figure}

In addition, it is worthy mentioning that the time tree structure is used to organize the temporal information associated with every hour during one or several days. Considering the fact that this graph based simulation platform is mainly used for training purpose, the trainees can easily modify and submit bidding data and other attribute parameters stored in graph edges of time tree to see the effect on market clearance and system operation. In Figure \ref{timeTree}, a time tree data structure with dashed circles is explicitly printed to show their connection relationship, in which every hour node is connected to bus node in power network and feeds corresponding temporal information respectively. However, the time tree structure is not necessary to be printed although it contains all the needed data and hides intentionally from the screen.

\subsection{IEEE-118 bus system and performance analysis}
\begin{figure}[h]
\centering
\includegraphics[width=0.48\textwidth]{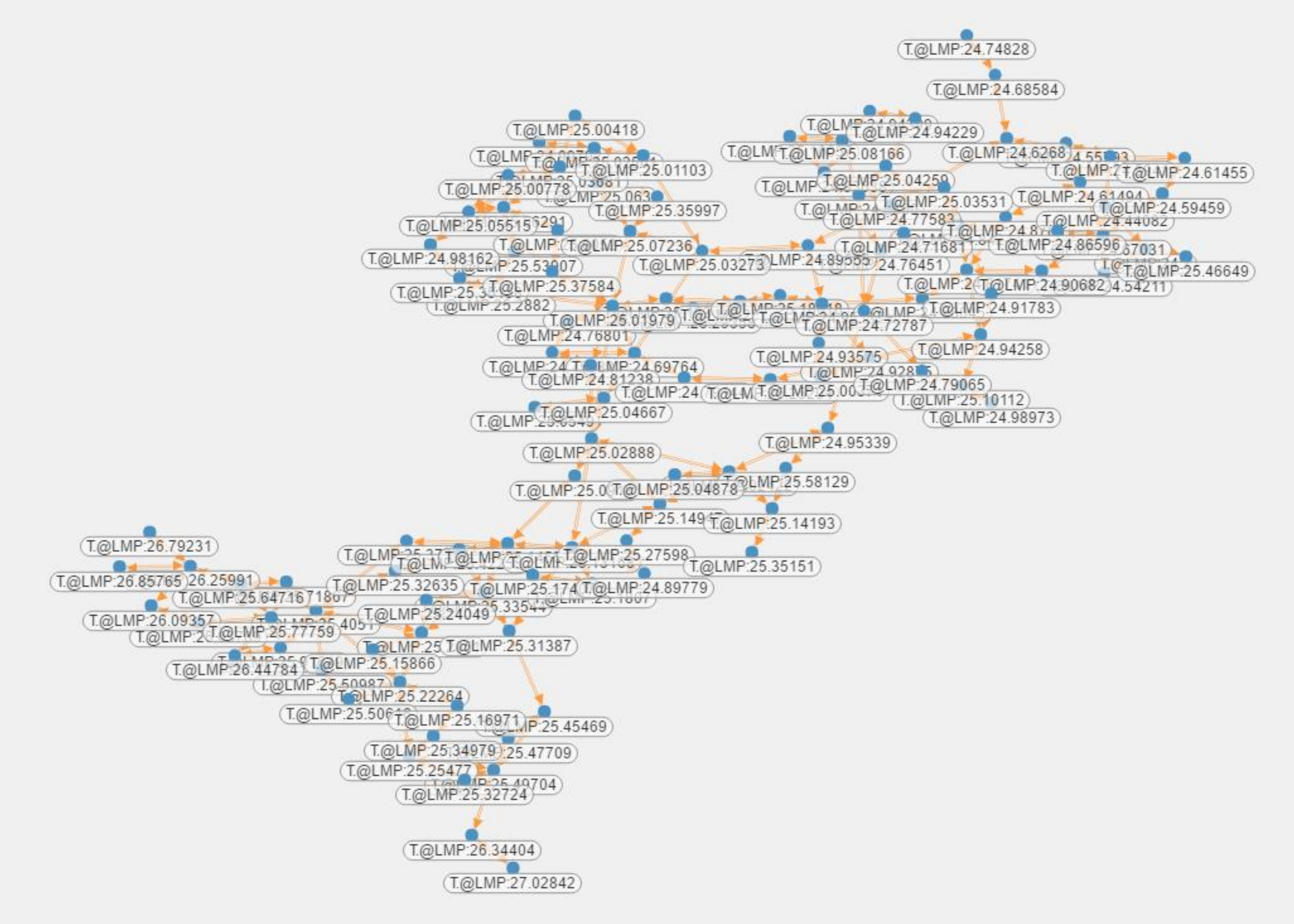}
\caption{118 bus system with visualization on GraphStudio}
\label{118bus}
\end{figure}

The IEEE-118 system has also been tested out as shown in Figure \ref{118bus} and used to observe the performance of parallel implementation within graph database system. Due to the page limit, some performance analysis can be found in other works of our group. The graph computing method can almost keep linear computational complexity rather than exponential increase of other methods.

\section{Conclusion}
Deregulated electricity market for more participants is a main a driven force for the further development of advanced smart grid technologies, especially for the power sector reform effort recently. In this paper, an innovative graph computing framework based on TigerGraph is proposed to facilitate the electricity market clearing process by utilizing graph parallelized power flow for SCUC and SCED problems. It provides many flexible interfaces for market or network data storage/operation with help of GSE/GPE  blocks in TigerGraph. Additionally, a comprehensive visualization platform based on GraphStudio is used to demonstrate the electricity market clearing results and transmission congestion effect. The whole system can also be used for electricity market operators' education and training purpose, helping promote future market deregulation.


%




\ifCLASSOPTIONcaptionsoff
  \newpage
\fi



%
\bibliographystyle{IEEEtran}

\bibliography{mybibfile}

%








\end{document}